\documentclass[runningheads,a4paper]{llncs}

\usepackage{amssymb}
\usepackage{graphicx}

\usepackage{url}
\urldef{\mailsa}\path| mahyunst@yahoo.com, mahyuddin@usu.ac.id|
\newcommand{\keywords}[1]{\par\addvspace\baselineskip
\noindent\keywordname\enspace\ignorespaces#1}

\begin{document}

\mainmatter  

\title{Algebraic on Magic Square of Odd Order $n$}

\titlerunning{Algebraic on Magic Square of Odd Order $n$}
 
\author{Mahyuddin K. M. Nasution}
\authorrunning{M. K. M. Nasution}
\institute{Mathematic Department, Fakultas Matematika dan Ilmu Pengetahuan Alam\\
Universitas Sumatera Utara, Padang Bulan 20150 USU Medan Indonesia\\
\mailsa\\}

\toctitle{Bulletin of Mathematics}
\tocauthor{}
\maketitle

\begin{abstract}
This paper aims to address the relation between a magic square of odd order $n$ and a group, and their properties. By the modulo number $n$, we construct entries for each table from initial table of magic square with large number $n^2$. Generalization of the underlying idea is presented, we obtain unique group, and we also prove variants of the main results for magic cubes.
\end{abstract}

\section{Introduction}

According to the book of W. S. Andrews \cite{andres1960}, the study of magic squares is quite old and dates back to ancient Tibet, to 12th century China, to 9th century Arab astrologers and perhaps much further. Speculation about it might even be prehistoric. 

In this paper, we shall see old procedure can product unique magic square based on a group of a set f numbers in modulo $n$. Objectives are find new magic square which it has different procedure if it compares with old. Therefore, we get a procedure for generating new magic square. Of course, this paper organized by first we defined the magic square with a condition for producting it, and based on conditions we result a procedure on odd order $n$. Next section, we modify all entries of magic square on modulo n and we test all conditions of magic square. In finding new procedure for the magic square with all entries on modulo $n$, we force a simple group on a set $Z_n$ with a binary operation, and based on we find new magic square. What are all magic squares of odd order n satisfying the condition of magic square? 

\section{The Magic Squares}

A magic square (MS) of order $n$, also called a $n\times n$ magic square, or it means that $n\times n$ square array  $A = (a_{ij})$ $0\leq$, $j\leq n-1$, of positive integers such that
\begin{itemize}
\item[a.] each integer from $1$ to $n^2$ inclusive occurs exactly once among the entries of $A$,
\item[b.] for $0\leq i\leq n-1$, the sum $\sum_{j=0}^{n-1}a_{ij}$ is independent of $i$,
\item[c.] for $0\leq i\leq n-1$, the sum $\sum_{j=0}^{n-1}a_{ij}$ is independent of $j$,
\item[d.] the sums $\sum_{i=0}^{n-1}a_{ii}$ and $\sum_{i=0}^{n-1} a_{i,n-i-1}$ are equal to the sums given in (b.), and such as to those in (c.).
\end{itemize}

\begin{center}
\begin{tabular}{|c|c|c|}\hline
8 & 1 & 6\cr\hline
3 & 5 & 7\cr\hline
4 & 9 & 2\cr\hline
\end{tabular}\\
{\bf Fig. 1.} $3\times 3$ magic square (of odd order).
\end{center}

\begin{center}
\begin{tabular}{|c|c|c|c|}\hline
1&15&14&4\cr\hline
12 &6&7&9\cr\hline
8&10&11&5\cr\hline
13&3&2&16\cr\hline
\end{tabular}\\
{\bf Fig. 1.} $4\times 4$ magic square (of even order).
\end{center} 

Let $A$ be the $3\times 3$ magic square for example in Fig. 1, and $B$ be the $4\times 4$ magic square in Fig. 2. The product square $A*B$ has order 12, the product of the orders of $A$ and $B$. Let $G$ be an abelian group and choose an element $u$ of $G$ once and for all. Denote by $A(G)$ the set of all square arrays of elements of $G$, the size of the arrays being arbitrary. If $A =(a_{ij})$, $0\leq i$, $j\leq m-1$, and $B = (b_{kl})$, $0\leq k$, $l\leq n-1$ are elements of $A(G)$ of size $m\times m$ and $n\times n$ respectively, then $A*B$ will be the $mn\times mn$ matrix $E = (e_{\alpha\beta})$, $(\alpha,\beta)$-the entry.

A set $S$ closed under an associative operation is called a semigroup, and the set be monoid if its operation with an identity element. The following are some results \cite{adler1997}. 

\begin{lemma}
Let $G$ be an abelian group. The set $A(G)$ of al square arrays with entries in $G$ is a monoid with identity element $u$ with respect to the operation $*$ defined by
\begin{equation}
e_{\alpha\beta} = m^2(b_{kl}+u)+a_{ij}
\end{equation}
and
\begin{equation}
(\alpha,\beta) = m(k,l)+(i,j)
\end{equation}
\end{lemma}

Let $S$ be a monoid with identity element of $S$ that respect to an operation. We say that $S$ is left cancelleative for the other side. In this case, the entries of a magic square of order $n$ run from 0 to $n^2-1$, and we take $u=0$ bukan $u=-1$.

For $A$ is $m\times m$, $B$ is $n\times n$ and $C$ is $p\times p$. If $A*C=B*C$ then we must have $mp=np$ and thus $m=n$. Therefore, we have $m^2(c_{kl}-1)+a_{ij} = m^2(c_{kl}-1)+b_{ij}$ for $0\leq i$, $j\leq m-1$ and $0\leq k$, $l\leq n-1$, which implies that $a_{ij}=b_{ij}$ for all $i$, $j$. That means that $A*B = B8C$, it implies $A=B$. If $A*B = A*C$ then we must have $mn=mp$ and therefore $n=p$. It follows that $m^2(b_{kl}+1)+a_{ij} = m^2(c_{kl}-1)+a_{ij}$ for $0\leq i$, $j\leq m-1$ and $0\leq k$, $l\leq n-1$, which implies that $m^2(b_{kl}-c_{kl} = 0$ for all $k$ and $l$. It follows that $b_{kl} = c_{kl}$ for all $k, l$, and equation $A*B=A*C$ implies $B=C$.

\begin{lemma}
Let $G$ be an abelian group and let $u$ be an element of $G$. Then the monoid $(A(G),*,u)$ is right cancellative.
\end{lemma} 

\begin{lemma}
Let $G$ be an abelian group and let $u$ be an element of $G$. Then the monoid $(A(G),*,u)$ is left cancellative if and only if the group $G$ is torsion-free.
\end{lemma}

\section{The Odd Order}

Let we consider the magic square of od order $n$, $n =3,5,7,9,\dots$. All condition (a.) - (b.) above give entries of a magic square of odd order $n$ based on following steps:

\begin{enumerate}
\item Set $i = (n-1)/2$, $j = 0$, and $k=1$.
\item Do while $k\leq n^2$
\begin{enumerate}
\item If $j=-1$ Then
\begin{enumerate}
\item If $i=n$ Then $i=i-1$ and $j=j+2$ Else $j=n-1$ 
\end{enumerate}
\item If $i=n$ Then $i=0$
\item If $a_{ij}\geq 0$ Then $j=j+2$ and $i=i-1$
\item $a_{ij}=k$
\item $j=j-1$, $i=i+1$ and $k=k+1$
\end{enumerate}
\item End Do
\end{enumerate}

For example, we obtain the sequential numbers from lines of $n\times n$ magic squares:\\
$n = 5$, it is $17,24,1,8,15,23,5,7,14,16,4,6,13,20,22,10,12,19,21,3,11,18,25,$\\
$2,9$; \\
$n = 7$, it is $30,39,48,1,10,19,28,38,47,7,9,18,27,29,46,6,8,17,26,35,37,5,14,$\\
$16,25,34,36,45,13,15,24,33,42,44,4,21,23,32,41,43,3,12,22,31,40,49,2,11,$\\
$20$; and\\
$n = 9$, it is $47,58,69,80,1,12,23,34,45,57,68,7,789,9,11,22,33,44,46,67,78,8,$\\
$10,21,32,43,54,56,77,7,18,20,31,42,53,55,66,6,17,19,30,41,52,63,65,76,16,$\\
$27,29,40,51,62,64,75,5,26,28,39,50,61,72,74,4,15,36,38,49,60,71,73,3,14,$\\
$25,37,48,59,70,81,2,13,24,35$;\\
are magic squares of odd ordern $n = 5,7,9$.

If we modify condition (a.) of magic square so that the entries of a magic square of odd order $n$ run from 1 to $n^2$ modulo $n$ instead of from 1 to $n^2$, we get a difference between a column (left side) and next column (to right side) is 2 in modulo $n$ and a difference betwen a row and next row from top to down is 1 in modulo $n$. Therefore, we define function $f : Z_n\times Z_n\rightarrow Z_n$ or a binary operation on $Z_n$ defined by $f(r,c) = 2+r+2c$, for all $r, c\in Z_n$, $r$ and $c$ represent row and column of magic square, respectively. For example, $n = 3,5,7$ and $9$, we have new squares likes Figs. 3, 4, 5 and 6. Each of this squares in $Z_n$ with $f$, $n = 3k$, $k=1,3,5,7,9,\dots$. We obtain result of condition (d.) of a magic square, $\sum_{i=0}^{n-1}$ (we called it as sltdd or \emph{sums of left-top-down-diagonal}) is not equal to $\sum_{i=0}^{n-1}$ (we called it as slbud or \emph{sums of left-buttom-up-diagonal}) or other conditions (sr = sums of row and sc = sums of column), see Table A in Appendix.

\begin{center}
\begin{tabular}{|c|c|c|}\hline
2 & 1 & 0\cr\hline
0 & 2 & 1\cr\hline
1 & 0 & 2\cr\hline
\end{tabular}\\
{\bf Fig. 3.} $3\times 3$ magic square in modulo $3$.
\end{center}

\begin{center}
\begin{tabular}{|c|c|c|c|c|}\hline
2 & 4 & 1 & 3 & 0\cr\hline
3 & 0 & 2 & 4 & 1\cr\hline
4 & 1 & 3 & 0 & 2\cr\hline
0 & 2 & 4 & 1 & 3\cr\hline
1 & 3 & 0 & 2 & 4\cr\hline
\end{tabular}\\
{\bf Fig. 4.} $5\times 5$ magic square in modulo $5$.
\end{center}

\begin{center}
\begin{tabular}{|c|c|c|c|c|c|c|}\hline
2 & 4 & 6 & 1 & 3 & 5 & 0\cr\hline
3 & 5 & 0 & 2 & 4 & 6 & 1\cr\hline
4 & 6 & 1 & 3 & 5 & 0 & 2\cr\hline
5 & 0 & 2 & 4 & 6 & 1 & 3\cr\hline
6 & 1 & 3 & 5 & 0 & 2 & 4\cr\hline
0 & 2 & 4 & 6 & 1 & 3 & 5\cr\hline
1 & 3 & 5 & 0 & 2 & 4 & 6\cr\hline
\end{tabular}\\
{\bf Fig. 5.} $7\times 7$ magic square in modulo $7$.
\end{center}

\begin{center}
\begin{tabular}{|c|c|c|c|c|c|c|c|c|}\hline
2 & 4 & 6 & 8 & 1 & 3 & 5 & 7 & 0\cr\hline
3 & 5 & 7 & 0 & 2 & 4 & 6 & 8 & 1\cr\hline
4 & 6 & 8 & 1 & 3 & 5 & 7 & 0 & 2\cr\hline
5 & 7 & 0 & 2 & 4 & 6 & 8 & 1 & 3\cr\hline
6 & 8 & 1 & 3 & 5 & 7 & 0 & 2 & 4\cr\hline
7 & 0 & 2 & 4 & 6 & 8 & 1 & 3 & 5\cr\hline
8 & 1 & 3 & 5 & 0 & 0 & 2 & 4 & 6\cr\hline
0 & 2 & 4 & 6 & 8 & 1 & 3 & 5 & 7\cr\hline
1 & 3 & 5 & 0 & 0 & 2 & 4 & 6 & 8\cr\hline
\end{tabular}\\
{\bf Fig. 6.} $9\times 9$ magic square in modulo $9$.
\end{center}

A set $Z_n$ with a binary operation $f$ is not an associative operation, and then $(Z_n,f)$ is not a group. Let us have a procedure to product new magic square (it is called New MS), may be, so that a set $Z_n$ with a binary operation satisfies all condition of group. The procedure as follows:

\begin{enumerate}
\item Copy $a_{n-1,j}$ to first column.
\item Copy $a_{(n-1)/2,j}$ to second column.
\item Copy $a_{0,j}$ to third column.
\item Set $i=3$, $x=0$, $y=0$
\item Do While $i<0$
\begin{enumerate}
\item If ($i$ modulo $2=0$) Then 
\begin{enumerate}
\item $x = x+1$
\item Copy $a_{xj}$ to $a_{ij}$
\end{enumerate}
\item Else
\begin{enumerate}
\item $y=y+1$
\item Copy $a_{(n-1)/2+y,j}$ to $a_{ij}$
\end{enumerate}
\end{enumerate}
\item End Do
\end{enumerate}

For example, for the entries of a pair of squares, we run $1$ to $n^2$ and $1$ to $n^2$ modulo $n$, $n = 3,5,7,9$, respectively, and we obtain some equivalent squares, Fig. 6, 7, 8, 9.

\begin{center}
\begin{tabular}{|c|c|c|}\hline
6 & 1 & 8\cr\hline
7 & 5 & 3\cr\hline
2 & 9 & 4\cr\hline
\end{tabular} $\rightarrow$ 
\begin{tabular}{|c|c|c|}\hline
0 & 1 & 2\cr\hline
1 & 2 & 0\cr\hline
2 & 0 & 1\cr\hline
\end{tabular}\\
{\bf Fig. 6.} $3\times 3$ magic square or square in modulo $3$.
\end{center}

\begin{center}
\begin{tabular}{|c|c|c|c|c|}\hline
15 & 1  & 17 & 8 & 24\cr\hline
16 & 7  & 23 & 14 & 5\cr\hline
22 & 13 & 4  & 20 & 6\cr\hline
3  & 19 & 10 & 21& 12\cr\hline
9 & 25 & 11 & 2 & 18\cr\hline
\end{tabular} $\rightarrow$ 
\begin{tabular}{|c|c|c|c|c|}\hline
0 & 1 & 2 & 3 & 4\cr\hline
1 & 2 & 3 & 4 & 0\cr\hline
2 & 3 & 4 & 0 & 1\cr\hline
3 & 4 & 0 & 1 & 2\cr\hline
4 & 0 & 1 & 2 & 3\cr\hline
\end{tabular}\\
{\bf Fig. 7.} $5\times 5$ magic square or square in modulo $5$.
\end{center}

\begin{center}
\begin{tabular}{|c|c|c|c|c|c|c|}\hline
28 & 1  & 30 & 10 & 39 & 19 & 48\cr\hline
29 & 9  & 38 & 18 & 47 & 27 & 7 \cr\hline
37 & 17 & 46 & 26 & 6  & 35 & 8\cr\hline
45 & 25 & 5  & 34 & 14 & 36 & 16\cr\hline
4  & 33 & 13 & 42 & 15 & 44 & 24\cr\hline
12 & 41 & 21 & 43 & 23 & 3 & 32\cr\hline
20 & 49 & 22 & 2 & 31 & 11 & 40\cr\hline
\end{tabular} $\rightarrow$ 
\begin{tabular}{|c|c|c|c|c|c|c|}\hline
0 & 1 & 2 & 3 & 4 & 5 & 6\cr\hline
1 & 2 & 3 & 4 & 5 & 6 & 0\cr\hline
2 & 3 & 4 & 5 & 6 & 0 & 1\cr\hline
3 & 4 & 5 & 6 & 0 & 1 & 2\cr\hline
4 & 5 & 6 & 0 & 1 & 2 & 3\cr\hline
5 & 6 & 0 & 1 & 2 & 3 & 4\cr\hline
6 & 0 & 1 & 2 & 3 & 4 & 5\cr\hline
\end{tabular}\\
{\bf Fig. 8.} $7\times 7$ magic square or square in modulo $7$.
\end{center}

\begin{center}
\begin{tabular}{|c|c|c|c|c|c|c|c|c|}\hline
45 & 1  & 47 & 12 & 58 & 23 & 69 & 34 & 80\cr\hline
46 & 11 & 57 & 22 & 68 & 33 & 79 & 34 & 9\cr\hline
56 & 21 & 67 & 32 & 78 & 43 & 8  & 54 & 10\cr\hline
66 & 31 & 77 & 42 & 7  & 53 & 18 & 55 & 20\cr\hline
76 & 41 & 6  & 52 & 17 & 63 & 19 & 65 & 30\cr\hline
5  & 51 & 16 & 62 & 27 & 64 & 29 & 75 & 40\cr\hline
15 & 61 & 26 & 72 & 28 & 74 & 39 & 4  & 50\cr\hline
25 & 71 & 36 & 73 & 38 & 3  & 49 & 14 & 60\cr\hline
35 & 81 & 37 & 2  & 48 & 13 & 59 & 24 & 70\cr\hline
\end{tabular} $\rightarrow$ 
\begin{tabular}{|c|c|c|c|c|c|c|c|c|}\hline
0 & 1 & 2 & 3 & 4 & 5 & 6 & 7 & 8\cr\hline
1 & 2 & 3 & 4 & 5 & 6 & 7 & 8 & 0\cr\hline
2 & 3 & 4 & 5 & 6 & 7 & 8 & 0 & 1\cr\hline
3 & 4 & 5 & 6 & 7 & 8 & 0 & 1 & 2\cr\hline
4 & 5 & 6 & 7 & 8 & 0 & 1 & 2 & 3\cr\hline
5 & 6 & 7 & 8 & 0 & 1 & 2 & 3 & 4\cr\hline
6 & 7 & 8 & 0 & 1 & 2 & 3 & 4 & 5\cr\hline
7 & 8 & 0 & 1 & 2 & 3 & 4 & 5 & 6\cr\hline
8 & 0 & 1 & 2 & 3 & 4 & 5 & 6 & 7\cr\hline
\end{tabular}\\
{\bf Fig. 9.} $9\times 9$ magic square or square in modulo $9$.
\end{center}

The squre of New MS represent a binary operation, $g : Z_n\times Z_n\rightarrow Z_n$, or $g(a,b) = a+b$ for example, for all $a,b\in Z_n$, and then we have exactly a set $Z_n$ with $g$ construct a simple group ($G$ mod $n$) where $\sum_{i=0}^{n-1} a_{i,n-i-1}$ is $2\sum_{i=0}^{n-1} a_{ii}$ or $2\sum_{j=0}^{n-1} a_{ij}$ ro $2\sum_{j=0}^{n-1} a_{ij}$. Each New MS of $n=5,7,11,\dots$ satisfies conditions (a.)-(d.) of magic square, except a part of condition (d.), i.e. $\sum_{i=0}^{n-1} a_{i,n-i-1}$, but for $n=3$ and $n=9$ two cases where all conditions of magic square are holded. Therefore, there exist two simple group of $Z_n$ with $g$, $n = 3,9$, we call them as \emph{pure magic square}. 

Let $M_2$ denote the set of all magic squares, include New MS of odd order $n=3$ and $n=9$. The magic square $A*B$ formed from $A$ is New MS $n=3$ and $B$ is $n=34$ and $n=4$, see next figures.

\begin{center}
\begin{tabular}{||c|c|c||c|c|c||c|c|c||c|c|c||}\hline\hline
51 & 46 & 53 &  6 &  1 &  8 & 69 & 64 & 71\cr\hline
52 & 50 & 48 &  7 &  5 &  3 & 70 & 68 & 66\cr\hline
47 & 54 & 49 &  2 &  9 &  4 & 65 & 72 & 67\cr\hline\hline
60 & 55 & 62 & 42 & 37 & 44 & 24 & 19 & 26\cr\hline 
61 & 59 & 57 & 43 & 41 & 39 & 25 & 23 & 21\cr\hline
56 & 63 & 58 & 38 & 45 & 40 & 20 & 27 & 22\cr\hline\hline
15 & 10 & 17 & 78 & 73 & 80 & 33 & 28 & 35\cr\hline
16 & 14 & 12 & 79 & 77 & 75 & 34 & 32 & 30\cr\hline
11 & 18 & 13 & 74 & 81 & 76 & 29 & 36 & 31\cr\hline\hline
\end{tabular}\\
{\bf Fig. 10.} $3\times 3$ magic square in $3\times 3$ magic square.
\end{center}

We locate the square in B  which contains the number 1 and place a copy of A in the corresponding square of the frame we have just constructed. Next we locate the square in B containing 2 and in the corresponding the square, we count out the next 9 numbers in the same pattern. It is the same to say that we adds 9 for all of the entries of A and places the result in the box corresponding to the position of the 2 in B. Next we find the 3 of B and we counts out the next 9 numbers in the corresponding place in the frame. Continuing in this way, we eventually get the magic square completely by this method.

\begin{center}
\begin{tabular}{||c|c|c||c|c|c||c|c|c||c|c|c||}\hline\hline
69 & 64 & 71 &  6 &  1 &  8 & 51 & 46 & 53\cr\hline
70 & 68 & 66 &  7 &  5 &  3 & 52 & 50 & 48\cr\hline
65 & 72 & 67 &  2 &  9 &  4 & 47 & 54 & 49\cr\hline\hline
24 & 19 & 26 & 42 & 37 & 44 & 60 & 55 & 62\cr\hline 
25 & 23 & 21 & 43 & 41 & 39 & 61 & 59 & 57\cr\hline
20 & 27 & 22 & 38 & 45 & 40 & 56 & 63 & 58\cr\hline\hline
33 & 28 & 35 & 78 & 73 & 80 & 15 & 10 & 17\cr\hline
34 & 32 & 30 & 79 & 77 & 75 & 16 & 14 & 12\cr\hline
29 & 36 & 31 & 74 & 81 & 76 & 11 & 18 & 13\cr\hline\hline
\end{tabular}\\
{\bf Fig. 10.} $3\times 3$ magic square in a modified $3\times 3$ magic square.
\end{center}

\begin{center}
\begin{tabular}{||c|c|c||c|c|c||c|c|c||c|c|c||}\hline\hline
53 & 46 & 51 &  8 &  1 &  6 & 71 & 64 & 69\cr\hline
48 & 50 & 52 &  3 &  5 &  7 & 66 & 68 & 70\cr\hline
49 & 54 & 47 &  4 &  9 &  2 & 67 & 72 & 65\cr\hline\hline
62 & 55 & 60 & 44 & 37 & 42 & 26 & 19 & 24\cr\hline 
57 & 59 & 61 & 39 & 41 & 43 & 21 & 23 & 25\cr\hline
58 & 63 & 56 & 40 & 45 & 38 & 22 & 27 & 20\cr\hline\hline
17 & 10 & 15 & 80 & 73 & 78 & 35 & 28 & 33\cr\hline
12 & 14 & 16 & 75 & 77 & 79 & 30 & 32 & 34\cr\hline
13 & 18 & 11 & 76 & 81 & 74 & 31 & 36 & 29\cr\hline\hline
\end{tabular}\\
{\bf Fig. 11.} The modified $3\times 3$ magic square in a $3\times 3$ magic square.
\end{center}

\begin{center}
\begin{tabular}{||c|c|c||c|c|c||c|c|c||c|c|c||c|c|c||}\hline\hline
  6 &   1 &   8 & 132 & 127 & 134 & 123 & 118 & 125 &  33 &  28 &  35\cr\hline
  7 &   5 &   3 & 133 & 131 & 129 & 124 & 122 & 120 &  34 &  32 &  30\cr\hline
  2 &   9 &   4 & 128 & 135 & 130 & 119 & 126 & 121 &  29 &  36 &  31\cr\hline\hline
105 & 100 & 107 &  51 &  46 &  53 &  60 &  55 &  62 &  78 &  73 &  80\cr\hline 
106 & 104 & 102 &  52 &  50 &  48 &  61 &  59 &  57 &  79 &  77 &  75\cr\hline
101 & 108 & 103 &  47 &  54 &  49 &  56 &  63 &  58 &  74 &  81 &  76\cr\hline\hline
 69 &  64 &  71 &  87 &  82 &  89 &  96 &  91 &  98 &  42 &  37 &  44\cr\hline
 70 &  68 &  66 &  88 &  86 &  84 &  97 &  95 &  93 &  43 &  41 &  39\cr\hline
 65 &  72 &  67 &  83 &  90 &  85 &  92 &  99 &  94 &  38 &  45 &  40\cr\hline\hline
114 & 109 & 116 &  24 &  19 &  26 &  15 &  10 &  17 & 141 & 136 & 143\cr\hline
115 & 113 & 111 &  25 &  23 &  21 &  16 &  14 &  12 & 142 & 140 & 138\cr\hline
110 & 117 & 112 &  27 &  22 &  11 &  18 &  18 &  13 & 137 & 144 & 139\cr\hline\hline
 \end{tabular}\\
{\bf Fig. 12.} The $3\times 3$ magic square in a $4\times 4$ magic square.
\end{center}

The product of two squares A*B be an order 12, it is the product of the orders of A and B. It is convenient to represent an analytic expression for this operation as (1) and (2) of Lemma 1. For n = 3 and 9 we can see easily that m = n, a binary operation (1) to be 
\[
e_{\alpha\beta} = n^2(b_{kl}+u)+a_{ij}
\]
For special case for $n=3$ and $n=9$ as new magic squares follow the Lemma 1, 2, and 3. 

\section{Conclusions}
In the simple group $Z_n$ we can product a procedure to construc a new magic square whereby they satisfies all conditions of magic square, mainly for n = 3 and 9, and these new magic squares so satisfy Lemma 1, 2 and 3.

\break
\noindent
{\bf Appendix}\\
\bigskip
\noindent
{\bf Table A.} Results of computation for all conditions of magic squares.\\
\begin{tabular}{|c|c|c|r|r|r|r|}\hline
$n$ & $n^2$ & Case & sr & sc & sltdd & slbud\cr\hline
3 & 9 & MS      & 15 & 15 & 15 & 15\cr
  &   & Mod $n$ & 3 & 3 & {\it 6} & 3\cr
  &   & $G$ mod $n$ & 3 & 3 & 3 & 6\cr
  &   & New MS  & 15 & 15 & 15 & 15\cr\hline
5 & 25& MS      & 65 & 65 & 65 & 65\cr
  &  & Mod $n$ & 10 & 10 & 10 & 10\cr
  &   & $G$ mod $n$ & 10 & 10 & 10 & 20\cr
  &   & New MS & 65 & 65 & 65 & {\it 70}\cr\hline
7 & 49& MS      & 175 & 175 & 175 & 175\cr
  &  & Mod $n$  & 21 & 21 & 21 & 21\cr
  &   & $G$ mod $n$ & 21 & 21 & 21 & 42\cr
  &   & New MS & 175 & 175 & 175 & {\it 189}\cr\hline
9 & 81& MS      & 369 & 369 & 369 & 369\cr
  &  & Mod $n$  & 36 & 36 & {\it 45} & 36\cr
  &   & $G$ mod $n$  & 36 & 36 & 36 & 72\cr
  &   & New MS & 369 & 369 & 369 & 369\cr\hline
11 & 121& MS      & 671 & 671 & 671 & 671\cr
  &  & Mod $n$    & 55 & 55 & 55 & 55\cr
  &   & $G$ mod $n$  & 55 & 55 & 55 & 110\cr
  &   & New MS & 671 & 671 & 671 & {\it 715}\cr\hline
\multicolumn{7}{|c|}{$\cdots$}\cr\hline
99 & 9801& MS      & 485199 & 485199 & 485199 & 485199\cr
  &  & Mod $n$    & 4851 & 4851 & 4851 & 4851\cr
  &   & $G$ mod $n$  & 4851 & 4851 & 4851 & 9702\cr
  &   & New MS & 485199 & 485199 & 485199 & {\it 489951}\cr\hline
\multicolumn{7}{|c|}{$\cdots$}\cr\hline
\end{tabular}
\end{document}